\documentclass[a4paper, amsfonts, amssymb, amsmath, pra, twocolumn, showkeys, nofootinbib, twoside]{revtex4}
\usepackage[english]{babel}
\usepackage[utf8]{inputenc}
\usepackage{amsmath}
\usepackage{bmpsize}
\usepackage{graphicx}
\usepackage{amsthm}
\usepackage{mathtools}
\usepackage{physics}
\usepackage{xcolor}
\usepackage[left=23mm,right=13mm,top=35mm,columnsep=15pt]{geometry} 
\usepackage{adjustbox}
\usepackage{placeins}
\usepackage[T1]{fontenc}
\usepackage{lipsum}
\usepackage{csquotes}
\usepackage{braket}

\usepackage[colorinlistoftodos, color=green!40, prependcaption]{todonotes}
\usepackage{graphicx}
\usepackage{amsthm}
\usepackage{mathtools}
\usepackage{physics}
\usepackage{xcolor}
\usepackage[left=23mm,right=13mm,top=35mm,columnsep=15pt]{geometry} 
\usepackage{adjustbox}
\usepackage{placeins}
\usepackage[T1]{fontenc}
\usepackage{lipsum}
\usepackage{csquotes}

\usepackage[pdftitle={Article}, pdfauthor={Author}]{hyperref} % For hyperlinks in the PDF
\begin{document}
\title{Quantum Steganography via Coherent and Fock State Encoding in an Optical Medium}
\author{Bruno Avritzer}
\email{avritzer@usc.edu}
\affiliation{Department of Physics and Astronomy, University of Southern California, Los Angeles, California}
\author{Todd A. Brun}
 \email{tbrun@usc.edu}
\affiliation{Ming Hsieh Department of Electrical and Computer Engineering, University of Southern California, Los Angeles, California}

\date{\today} % Leave empty to omit a date

\begin{abstract}
Steganography is an alternative to cryptography, where information is protected by secrecy---being disguised as innocent communication or noise---rather than being scrambled. In this work we develop schemes for steganographic communication using Fock and coherent states in optical channels based on disguising the communications as thermal noise, with a fidelity approaching 1. We derive bounds on their efficiency in terms of the communication rate and the required keyrate in the case of an all-powerful eavesdropper, and provide explicit methods of encoding and error correction for the noiseless channel case. 
\end{abstract}

\keywords{quantum steganography, quantum communication, quantum information, covert communication}

\maketitle

\section{Introduction} \label{sec:outline}

Covert communication is often associated with military or espionage applications, but some of the earliest applications were for personal use. In Ancient Egypt, hieroglyphs were used by royal scribes to send the pharaoh's messages covertly. In Roman times, the Caesar cipher, where each letter in a message was ``shifted'' by a predetermined amount (akin to a secret key), was devised \cite{caes}, this approach being refined into the Vigenere encoding around the 15th Century \cite{brit}. Over time, developments became more mathematically and technologically advanced. Famously, the use of the Enigma machine in World War 2 enabled the Germans to communicate covertly until the cipher was cracked by the Allied effort, giving them a strategic information advantage for the rest of the war \cite{Enig2,cryptorigin}.

The Allies were able to replicate, understand, and reproduce the technology of the Enigma machine in order to crack its cipher; thus the practical application of quantum computers and quantum information devices, which represent an additional level of technological sophistication, are evident. The question we would like to explore is how to leverage quantum computers and quantum information, a present and future technological advantage, to communicate covertly in a variety of situations.

The main scenario we will consider is one where the eavesdropper has access to the full contents of the messages being transmitted, but is made to believe they are innocuous, unlike in cryptography where the encrypted text is often nonsensical without a key and so arouses suspicion. This reflects a difference in the situation: in cryptography, the messages are often read secretly, while in steganography the eavesdropper is not secret and maintains everything out in the open. In many ways, this is advantageous---consider the wartime example. If the goal is not to arouse suspicion, it can be markedly more suspicious and dangerous to send messages of gibberish through wartime censors than to have a chat with your friend on the phone about normal topics---only this chat contains some hidden information. Other examples of steganography are an invisible watermark that can only be revealed with a procedure no one would think to do spontaneously \cite{cryptorigin}, or a secret encoding of information in an audio file \cite{audiosteg}. Some steganographic encodings are readable by anyone who thinks to look for them, relying on the concealment of the covering message, while others may require a shared secret key between the sender and receiver, just like many cryptosystems. Steganography can even be combined with cryptography as they are effectively independent measures---for example, by using any methods we describe in this paper to transmit an already cryptographically encoded message. This will, in general, require more shared secret key.

A large amount of work has been done on quantum cryptography \cite{source1, source2, source3}. The field of quantum steganography (and, broadly, covert communication) is smaller, but also includes a substantial body of relevant theoretical work. It has recently been shown that over $n$ uses of additive white Gaussian noise (AWGN), a number of bits proportional to $\sqrt{n}$ can be communicated covertly \cite{SRL} which was later generalized in  \cite{CovComGain}. Furthermore, a number of methods have been devised for such communications using quantum systems, as in \cite{qstegref1,qstegref2,bloch}. In this paper we will study the encoding of information in quantum states transmitted over an optical channel in such a way that it imitates thermal noise. This follows the broad approach of Brun and Shaw in \cite{SB1,SB2} (for qubit channels), and has been studied for optical channels in \cite{Wu}. Like the latter, our work follows the ``secrecy'' approach typical of steganography, in which the message is protected by the fact that its existence is concealed. This is as opposed to the ``security'' approach of standard cryptography, as well as methods such as spread spectrum and chaotic communication that are not generally secret at an information-theoretic standard \cite{chaos, spreadspec}. Steganography, as studied in this paper, provides formal guarantees of secrecy based on metrics of fidelity and trace distance, unlike the aforementioned approaches, while also functioning in narrow-band. Compared to \cite{Wu}, our work is experimentally simpler, though its practical performance at scale remains to be demonstrated. It also does not require any assumptions about the ability of the eavesdropper to detect the noise beyond their expectation of a thermal state, which enables a potentially greater ability to communicate (as quantified by the communication rate and rate of secret key consumption). This current work only treats classical communication, but it shows the kind of methods that could be used in future quantum steganographic encodings for entanglement distribution or quantum communication, perhaps drawing on techniques similar to those in \cite{Zhuang}. In this work, we prove secrecy by calculating the trace distance or fidelity between the ``innocent'' (thermal) state and the average state containing hidden information. This approach is sufficient to demonstrate the effectiveness of a steganographic method, and is simpler than the proofs needed to show security in cryptography.

In Sections ~\ref{sec:metrics} and ~\ref{sec:develop} we develop the machinery required to understand the communication process; in Section ~\ref{sec:mappings} we develop and analyze some simple encodings; and in Sections~\ref{sec:simulation} and ~\ref{sec:encodings} we do a more detailed analysis of their implementation and efficiency. Section ~\ref{sec:discussion} summarizes the results and discusses future work.
\section{Important Measures for Steganography} \label{sec:metrics}
In steganographic protocols such as those we will propose in later sections, there is a key trade-off that must be taken into account: that of the effective communication rate achievable in the channel as opposed to the similarity of the targeted ``innocent'' state with the actual channel state. In some schemes, such as the Fock encoding we will discuss later, these can be quantified by just two measures: the communication rate (which is given by $R=1+p_{\text{err}}\log p_{\text{err}}+(1-p_{\text{err}})\log(1-p_{\text{err}})$ for a binary channel with a probability $p_{\text{err}}$ of mistaking one symbol for another) and the trace distance between the channel state $\rho$ containing hidden information, represented by the density operator corresponding to the state of the channel over which information is being transmitted, and the ``innocent'' thermal state $\rho_{th}$. This trace distance is given by 
\begin{equation}
    D(\rho, \rho_{th})=\frac{1}{2}||\rho-\rho_{th}||
\end{equation}
where the above norm is the trace norm. Alternatively, one can use the state fidelity
\begin{equation}
    F(\rho, \rho_{th}) = \left(\text{Tr}\left(\sqrt{\sqrt{\rho}\rho_{th}\sqrt{\rho}}\right)\right)^2
\end{equation} as another measure of distance that functions similarly to the trace norm, although it is not a metric on the set of density matrices in the formal sense. The trace distance can be used to directly compute the minimum probability of mistaking $\rho$ for $\rho_{th}$ using a positive operator-valued measurement (POVM), and is therefore conceptually useful as a representation of secrecy. However, the fidelity is often easier to calculate, and can be used to bound the trace distance \cite{fvdg} --- it also can be interpreted as a probability of mistaking one state for another when at least one of the states is pure. In particular, if the fidelity is 1 (or approaching 1), the trace distance is 0 (or approaching 0) and the two states cannot be distinguished by any measurement.  

In schemes where additional practical constraints are imposed to facilitate communication (for example, reducing the choice of possible states used in encoding to make distinguishing them easier, as in some of the coherent state protocols discussed below), we can assume that the sender and receiver draw on a pre-shared secret key, unknown to the eavesdropper. This key could be, for example, a secret string of random bits. This key usage can be quantified by a secret key rate $K$, the number of bits of secret key consumed per channel use. The ratio $R/K$ or difference $R-K$ of the rates give additional measures of the usefulness of the scheme for steganographic communication.    
  
These measures can help us evaluate the effectiveness of different potential protocols. On one extreme, if Alice and Bob simply send the ``innocent'' state at every time interval (i.e. with probability 1), the fidelity metric will have its highest possible value, which is 1. However, this encoding has no communication rate, $R=0$. On the other hand, if Alice and Bob use a naive encoding that maps the input bits 0 and 1 to a fixed pair of orthogonal states, it will generally be impossible to have good fidelity with the thermal state, and Eve can easily detect that communication is happening. The goal of a good steganographic protocol is to encode the message---a string of input bits---into a sequence of states, such that after averaging over all possible messages (and also the pre-shared secret key, if any), the fidelity with a string of thermal states is close to 1, but Bob can also retrieve the encoded string with high probability.

The protocols we will consider deal with cases where the fidelity (or trace distance) is very close to 1, at least in an asymptotic sense; the communication rate is nonzero; and the system can be implemented physically via the transmission of physically realizable states. The first example we will discuss is a protocol using coherent states.

\section{Disguising coherent states as thermal noise in a channel} \label{sec:develop}

A coherent state is a state of a quantum oscillator or field mode. It is defined, for some complex $\alpha$, as 
\begin{equation}
    \ket{\alpha}=\sum_{n=0}^\infty \frac{\alpha^ne^{-|\alpha|^2/2}}{\sqrt{n!}}\ket{n} ,
\end{equation}
which is the result of acting with the displacement operator $D(\alpha)=e^{\alpha a^\dagger-\alpha^*a}$ on the $\ket{0}$ state of a harmonic oscillator. A thermal state is given for the same type of system by
\begin{equation}
    \rho_{th} = \frac{1}{Z} \sum_{n=0}^\infty e^{-\frac{\hbar\omega (n+1/2)}{k_BT}}\ket{n}\bra{n} ,
\end{equation}
where 
\begin{equation}
    Z=\sum_{n=0}^\infty e^{-\frac{\hbar\omega(n+1/2)}{k_BT}}=\frac{1}{2}\text{csch}\left(\frac{\hbar\omega}{2k_BT}\right)
\end{equation} is the partition function. 

If we describe the thermal state of a mode in a channel in terms of the average number of photons transmitted, known as
\begin{equation}
    \Bar{n}=\left(e^{\frac{\hbar\omega}{k_BT}}-1\right)^{-1} ,
\end{equation}
we can reformulate the expression for $\rho_{th}$ in a simpler way:
\begin{equation}
    \rho_{th}=\frac{1}{\Bar{n}+1}\sum_{n=0}^\infty \left(\frac{\Bar{n}}{\Bar{n}+1}\right)^n\ket{n}\bra{n} .
\end{equation}

We want to represent this using coherent states over the phase space described by $\ket{\alpha}=\ket{re^{i\theta}}$, as in the Glauber P-Representation \cite{Vourdas}:
\begin{equation}
    \begin{aligned}
      \rho_{th}&=\frac{1}{N}\int d^2\alpha e^{-c|\alpha|^2}\ket{\alpha}\bra{\alpha}\\&=\frac{1}{N}\int_0^\infty rdr\int_0^{2\pi} d\theta e^{-cr^2}\ket{re^{i\theta}}\bra{re^{i\theta}}\\
      &=\frac{2\pi}{N}\sum_{n=0}^\infty \frac{1}{n!}\int_0^\infty r^{2n+1}e^{-(c+1)r^2} \ket{n}\bra{n}\\
      &=\frac{\pi}{N}\sum_{n=0}^\infty \frac{1}{(c+1)^{n+1}}\ket{n}\bra{n},\\
      \end{aligned}
\end{equation}
\begin{equation}
    \begin{aligned}
       \implies \frac{\pi}{N(c+1)^{n+1}}=\frac{1}{\Bar{n}}\left(\frac{\Bar{n}}{\Bar{n}+1}\right)^{n+1},
    \end{aligned}
\end{equation}
\begin{equation}
    \implies c=\frac{1}{\Bar{n}}, N=\pi\Bar{n},
\end{equation}
\begin{equation}
    \implies \rho_{th}=\frac{1}{\pi\Bar{n}}\int_0^\infty dr\int_0^{2\pi} d\theta re^{-\frac{r^2}{\Bar{n}}}\ket{re^{i\theta}}\bra{re^{i\theta}}.
\end{equation}
We can integrate over $\theta$ and consider this as a probability distribution over coherent states $\ket{r}$ with $p(r)=\frac{2}{\Bar{n}}re^{-\frac{r^2}{\Bar{n}}}$, i.e., a Rayleigh distribution:
\begin{equation}
\mathrm{Rayleigh}\left(r; \sqrt{\frac{\Bar{n}}{2}}\right) = \frac{2}{\Bar{n}}re^{-\frac{r^2}{\Bar{n}}} .
\end{equation}
The median of this distribution is given by $r_{1/2}=\sqrt{\Bar{n}\ln2}$, which is a convenient point of separation if we want to send binary messages.

Because the set of coherent states is over-complete, the existence of a coherent-state representation is guaranteed. A related question is how well a set of coherent states with a set of $M$ randomly-chosen radii $\{r_j\}$ and uniformly random phases can approximate a thermal state. This gives a mixture \begin{equation}
    \rho_c=\frac{1}{M}\sum_{j=1}^Me^{-r_j^2}\sum_n\frac{r_j^{2n}}{n!}\ket{n}\bra{n} .
\end{equation}
The answer is remarkably simple:
\begin{equation}
\begin{aligned}
   \sqrt{F(\rho_{th}, \rho_c)}&\geq 1-\frac{\Bar{n}}{2M} \\
   \implies ||\rho_{th}-\rho_c||&\leq\sqrt{\frac{4\Bar{n}}{M}-\frac{\Bar{n}^2}{M^2}} ,
   \label{equation:fidelity}
\end{aligned}
\end{equation}
where $F$ is the average fidelity $\sqrt{F(\rho_{th},\rho_c)}=\text{Tr}\left(\sqrt{\rho_{th}^{1/2}\rho_c\rho_{th}^{1/2}}\right)$. This is relevant for ``Pairwise'' protocols we will discuss later.

Another important bound is on the same kind of setup, but without averaging over $\theta$. If we instead discretize the circle over $\theta$---that is, we consider a set of states $\ket{\alpha_{jk}}=\ket{r_je^{\frac{2\pi ik}{L}}}$ for sufficiently large $L$---we can do at least as well as the above result. A proof of both these bounds is contained in the appendix.

\section{Mappings for Quantum Steganography} \label{sec:mappings}

In this work we consider four main approaches to encoding information steganographically as states of light: the Fock state encoding, which requires no shared key; an encoding in coherent states without shared key; and two other encodings into coherent states that do require shared key: the Vertical Angles encoding and the Redefined Rayleigh Distribution encoding.

\subsection{Fock State Methods}

The scheme for Fock state methods is straightforward and has one clear advantage: Fock states are more easily distinguishable than coherent states, so the communication capacity is higher, although the problem of realizing an arbitrary Fock state in an experimental setting is also more challenging than for coherent states. The protocol we will describe requires only the preparation of multiphoton Fock states, and may be done using probabilistic operations such as photon addition. As such, it can in principle be done using only single photon sources, single photon detectors, and beam splitters, although when $\Bar{n}$ is high this may become experimentally difficult due to low observation probability. It should be noted that in that regime, the coherent state methods described in this paper perform more competitively with the Fock state methods due to the greater ease of distinguishing between different coherent states at a higher amplitude. This method, however, requires an encoding system from binary digits to Fock states which depends on the value of $\Bar{n}$, and is described in detail in Section~\ref{sec:encodings} and the Appendix. If we are dealing with Fock states in a noiseless channel, the problem is one of translating regular binary strings into binary strings with a certain number of 1s and 0s determined by $\Bar{n}$.  It is worth noting as an experimental consideration that, in cases where $\Bar{n}$ is low, the weight of the encoded text will be low on average and a sophisticated encoding may not be necessary. In cases where $\Bar{n}$ is very large, that may not be the case, and it would require us to define different Fock states as encoding the binary 1 or 0, according to the Boltzmann weights of such Fock states (for example, we might define all states below a certain value of $n$ as belonging to 0, and the others to belong to 1). In Section~\ref{sec:encodings} we will consider a more intermediate case for communication of a message of length $N$ bits and derive bounds on its efficiency.

\subsection{Distribution Coherent State Methods}

Since Equation \ref{equation:fidelity} describes the average fidelity, it makes sense to examine different approaches to  optimize this quantity, taking into account the non-orthogonality of the coherent states being measured. In all cases we will draw from distributions defined by
\begin{equation}
    \begin{aligned}
    \begin{cases}
    \rho_0(r)=\frac{2}{\Bar{n}}re^{-\frac{r^2}{\Bar{n}}}, & 0<r<r_{\frac{1}{2}},\\
    \rho_1(r)=\frac{2}{\Bar{n}}re^{-\frac{r^2}{\Bar{n}}}, & r_{\frac{1}{2}}<r<\infty,
    \end{cases}
    \end{aligned}
\end{equation}
representing the transmission of 0 and 1 from the sender. 

First, we consider the ``distribution'' case, which does not require key, where Alice draws a coherent state randomly from $\rho_0$ or $\rho_1$ and Bob has to try and guess which distribution it came from. Bob's ability to do this is bounded by the trace distance between the states $\rho_0$ and $\rho_1$. This is difficult to evaluate, but can be evaluated in terms of the cumulative distribution function of the Poisson processes, 
\begin{equation}
    Q_n=2^{-(\Bar{n}+1)}\sum_{k=0}^n\frac{(cr_{1/2}^2)^k}{k!}
\end{equation} for a process with $\lambda=(\Bar{n}+1)\ln2=cr_{1/2}^2$ and $\Tilde{Q}_n$ for $\lambda=\Bar{n}\ln2$, with $N_{\frac{1}{2}}$ the median of the Poisson process
(a full derivation is included in the appendix):
\begin{equation}
\begin{aligned}
  \frac{1}{2}\Big|\Big|\rho_0-\rho_1\Big|\Big|&=\frac{1}{2(n+1)}\sum_{n=0}^\infty \left(\frac{\Bar{n}}{\Bar{n}+1}\right)^n|1-2Q_n|\\&=2\Bigg(1+(2Q_{N_{\frac{1}{2}}}-1)\left(\frac{\Bar{n}}{\Bar{n}+1}\right)^{N_\frac{1}{2}+1}\\&-\Tilde{Q}_{N_\frac{1}{2}}\Bigg).
\end{aligned}
\end{equation}
From this we can calculate the probability of error $p_{\text{err}}$ and the communication rate $R$. Since the Poisson process is discrete, this curve has some kinks when $N$ increases, as seen in figure \ref{fig:vertical}. When $\Bar{n}$ is large, the states $\rho_0$ and $\rho_1$ are almost orthogonal; but when $\Bar{n}$ is $O(1)$ or smaller, the two states have significant overlap and are not perfectly distinguishable. In this case error correction may be necessary, and matching the thermal state may still require shared secret key, as discussed in Section~\ref{sec:encodings}. We consider two variations of this idea below.

\subsection{Pairwise Coherent State Methods}

A related method is to partition the thermal state as above and select a finite set of states from each half to send. As we showed above, this finite set approximates the thermal state very well and can be distinguished more easily from each other. This method may be easier to implement practically, as only sampling from a subset of $\rho_{th}$ is sufficient, but requires more secret key than the distribution methods, which will be quantified in Section~\ref{sec:encodings}.

\subsection{Vertical Angles}

\begin{figure*}
    \centering
    \includegraphics[clip,width=1.1\textwidth]{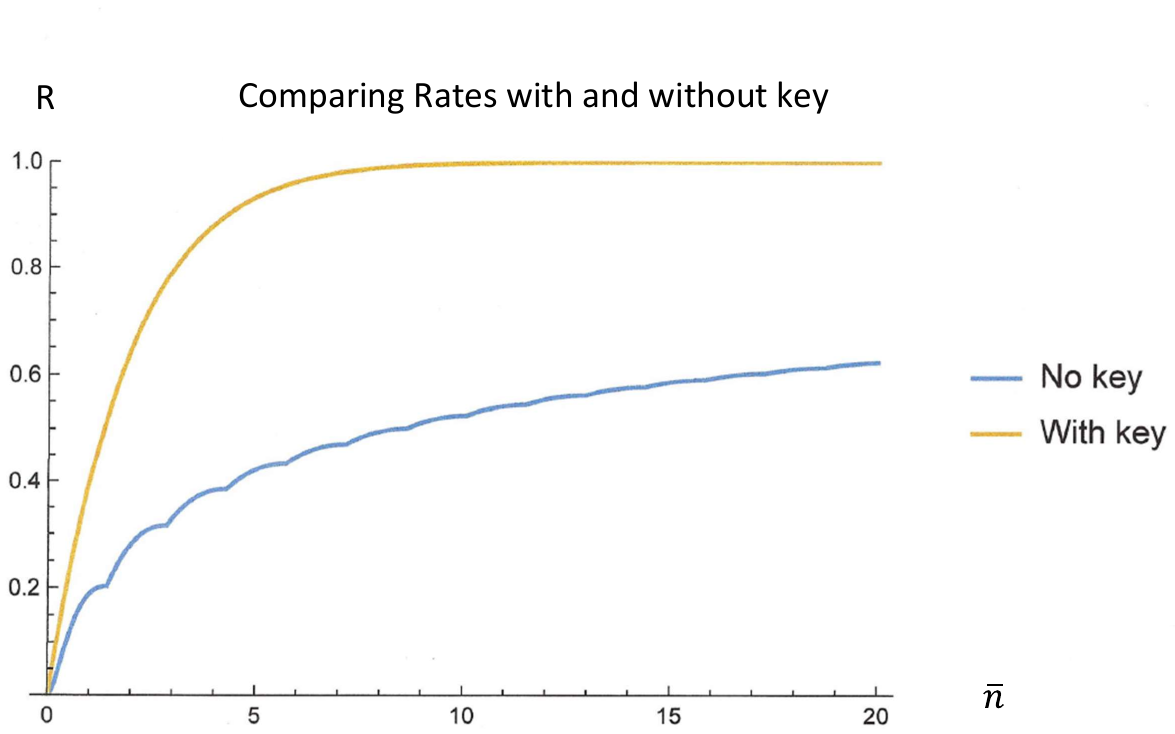}
    \caption{A plot of the lower bound on the communication rate for the vertical angle (key) encoding scheme compared to the ``Distribution'' (no key) scheme.}
    \label{fig:vertical}
\end{figure*}

In this approach, Alice and Bob choose $\alpha_0$ and $\alpha_1$ ahead of time to have opposite phases $\theta$, i.e. $\alpha_0=r_0e^{i\theta}$ and $\alpha_1=r_1e^{i(\theta+\pi)}=-r_1e^{i\theta}$, where $\theta$ can be chosen arbitrarily. These states correspond to the binary 0 or 1 and are drawn from the distributions $\rho_0$ and $\rho_1$, although since this specific protocol is more useful for the low-$\Bar{n}$ case, $\rho_0$ will likely be sampled more often. Knowledge of the two possible states constitutes a secret key, and the $\theta$ correlation helps distinguish the distributions by minimizing the overlap
\[
|\braket{\alpha_0}{\alpha_1}|=e^{-\frac{|\alpha_0-\alpha_1|^2}{2}}=e^{-\frac{|r_0+r_1|^2}{2}} . 
\]
This is maximized for $r_0=0$, $r_1=r_{\frac{1}{2}}$ which gives 
\begin{equation}
    |\braket{\alpha_0}{\alpha_1}|\leq e^{-\frac{r_{1/2}^2}{2}}=2^{-\frac{\Bar{n}}{2}},
\end{equation}
\begin{equation}
\begin{aligned}
     \implies p_{\text{err}}&=\frac{1}{2}\left(1-\Big|\Big|\ket{\alpha_0}\bra{\alpha_0}-\ket{\alpha_1}\bra{\alpha_1}\Big|\Big|\right)\\
     &=\frac{1}{2}\left(1-\sqrt{1-|\braket{\alpha_0|\alpha_1}|^2}\right)\\&\leq \frac{1}{2}\left(1-\sqrt{1-2^{-\Bar{n}}}\right).
\end{aligned}
\end{equation}
The associated communication rate is $R=1+p_{\text{err}}\log_2(p_{\text{err}})+(1-p_{\text{err}})\log_2(1-p_{\text{err}})$ which can be seen in figure~\ref{fig:vertical}.

We can think of this setup as using a key, since Alice and Bob must have prior information linking the two $\alpha$ values, in contrast to the aforementioned no-key case. Using this key allows a greater rate of communication, since the overlap between the coherent states representing 0 and 1 is minimized. We can see this in figure~\ref{fig:vertical}.

\subsection{Redefined Rayleigh Distributions}

The final approach involves simply drawing $r_0$ and $r_1$ from the corresponding distributions, communicating them over a channel, and attempting to determine which distribution was sampled from based on the channel measurement. This approach is equivalent to determining the overlap of the distributions $\rho_0$ and $\rho_1$, where we are once again randomizing $\theta$, but the two are not simultaneously diagonal in any basis and so the optimal measurement to distinguish them is not easy to find with no key. In this section we consider a specific type of measurement and attempt to optimize for a variable parameter $m_c$ which will distinguish between the two states $\rho_0$ and $\rho_1$, although in Section~\ref{sec:encodings} we will also consider measurements that saturate the Helstrom bound \cite{Helstrom}
\begin{equation}
    p_{\text{err}}\geq \frac{1}{2}-\frac{1}{2}||(1-f)\rho_0-f\rho_1||
\end{equation}
when discriminating between two states $\rho_0$ and $\rho_1$ occurring with probabilities $1-f$ and $f$, respectively.

There is one notable difference between this approach and the previous. For easy distinction, we can without loss of generality choose $\theta=0$ for our analysis and redefine the distribution $\Tilde{\rho}_1$ as spanning $(-\infty, -r_{1/2}]$ to make it easier to distinguish from $\Tilde{\rho}_0$, as this essentially rotates $\rho_1$ about the origin by $\pi$ and creates the greatest possible distance between the means and medians of $\rho_0$ and $\rho_1$ by such an operation, while not fundamentally changing the nature of any calculations we will perform.

\subsubsection{Setup}

\begin{figure}
    \centering
    \includegraphics[scale=.35]{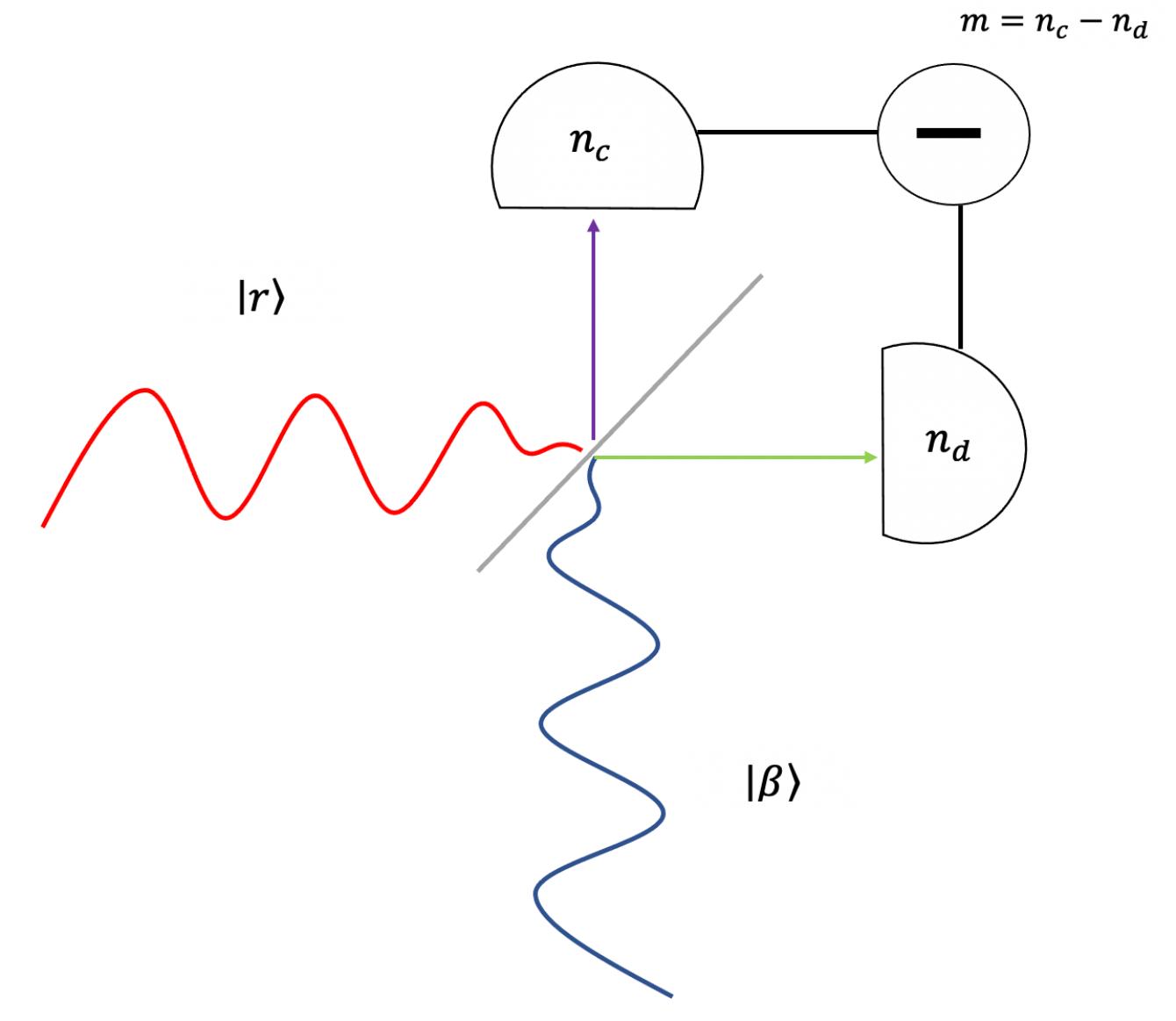}
    \caption{in this setup, a beam splitter combines the coherent states denoted by $\ket{r}$ and $\ket{\beta}$, with $n_c$ and $n_d$ denoting detectors that measure the incidence of photons. The value of the homodyne measurement is given by $m=n_c-n_d$.}
    \label{fig:setup}
\end{figure}

The coherent state-based protocols we are considering use a balanced homodyne measurement of a state $\ket{r}$. This state is coupled to an oscillator by means of the beam splitter shown in figure \ref{fig:setup}. If we have the operators $a$ and $a^\dagger$, of which $\ket{r}$ is an eigenket of $a$, and likewise $b$ and $b^\dagger$ for $\ket{\beta}$, once the states pass through the beam splitter the outcome is characterized by the new operators $c=\frac{a+b}{\sqrt{2}}$ and $d=\frac{a-b}{\sqrt{2}}$. As such, the desired observable is given by $m=n_c-n_d=c^\dagger c-d^\dagger d=a^\dagger b+b^\dagger a$.

\subsubsection{Bounds}

We can derive bounds using the generalized Markov inequality, 
\begin{equation}
    P(X-\mu>\lambda)\leq \frac{M_n(X)}{\lambda^n}
\end{equation}
for even $n$, where $X$ is a random variable, $\mu$ is its mean, and $M_n(X)$ is its $n$th moment.  

Applying this to the distributions for $r_1$ and $r_0$, we have for a cutoff value $m_c\in [2\beta\Bar{r}_1, 2\beta\Bar{r}_0]$ that %(see appendix for proof)
\begin{equation}
    \begin{aligned}
    P(m-\Bar{m_1}>\lambda)\leq \frac{M_n(m_1)}{\lambda^n},\\P(m-\Bar{m_0}<\lambda)\leq \frac{M_n(m_0)}{\lambda^n},
    \end{aligned}
\end{equation}
where here $\Bar{m}_1$ and $\Bar{m}_0$ refer to the mean expected $m$ values for each distribution, $2\beta\Bar{r}_1$ and $2\beta\Bar{r}_0$, respectively.

There is an explicit formula:
\begin{equation}
    M_n(X)=\sum_{k=0}^n\binom{n}{k}(-1)^{n-k}E[X^k](E[X])^{n-k}.
\end{equation}

We also have the explicit formula for moments of $m$, under the approximation $b\approx \beta$ (since only the leading order terms in $\beta$ matter for sufficiently large real $\beta$):
\begin{equation}
\begin{aligned}
  E(m^k)&=\int p(r)\beta^k\braket{r|(a+a^\dagger)^k|r}dr\\&=\frac{4\beta^k}{\Bar{n}}\int dr (re^{-\frac{r^2}{\Bar{n}}})\mathcal{F}_k(r)\text{ , }\\ 
  \end{aligned}
  \end{equation} where\begin{equation}
      \begin{aligned}
  \mathcal{F}_k(r)=\begin{cases}
  \sum_{j=0}^{k/2}\frac{k!2^{3j-k/2}}{(2j)!(k/2-j)!}r^{2j}, &\text{     k even}\\
  \sum_{j=0}^{(k-1)/2}\frac{k!2^{3j-\frac{k-3}{2}}}{(2j+1)!(\frac{k-1}{2}-j)!}r^{2j+1}, & \text{     k odd}.
  \end{cases}
\end{aligned}
\end{equation}
A derivation of the above is obtained by acting with $\braket{r|D(\alpha)}{r}=e^{2i\beta r-\frac{\beta^2}{2}}$ and equating the Taylor expansions of both sides to each order in $\beta$. This provides a means to evaluate the higher-moment Markov bound
\begin{equation}
\begin{aligned}
    p_{err}&=\frac{1}{2}[p(m-\Bar{m}>m_c-\Bar{m}|1)+p(m-\Bar{m}<m_c-\Bar{m}|0)]\\&\leq\frac{1}{2}\left[\frac{M_n(m_1)}{(m_c-\Bar{m}_1)^n}+\frac{M_n(m_0)}{(m_c-\Bar{m}_0)^n}\right].
\end{aligned}
\end{equation}

\section{Numerical simulation and performance} \label{sec:simulation}

\begin{figure*}
    \centering
    \includegraphics[trim = 50 300 0 200,clip, width=.8\textwidth]{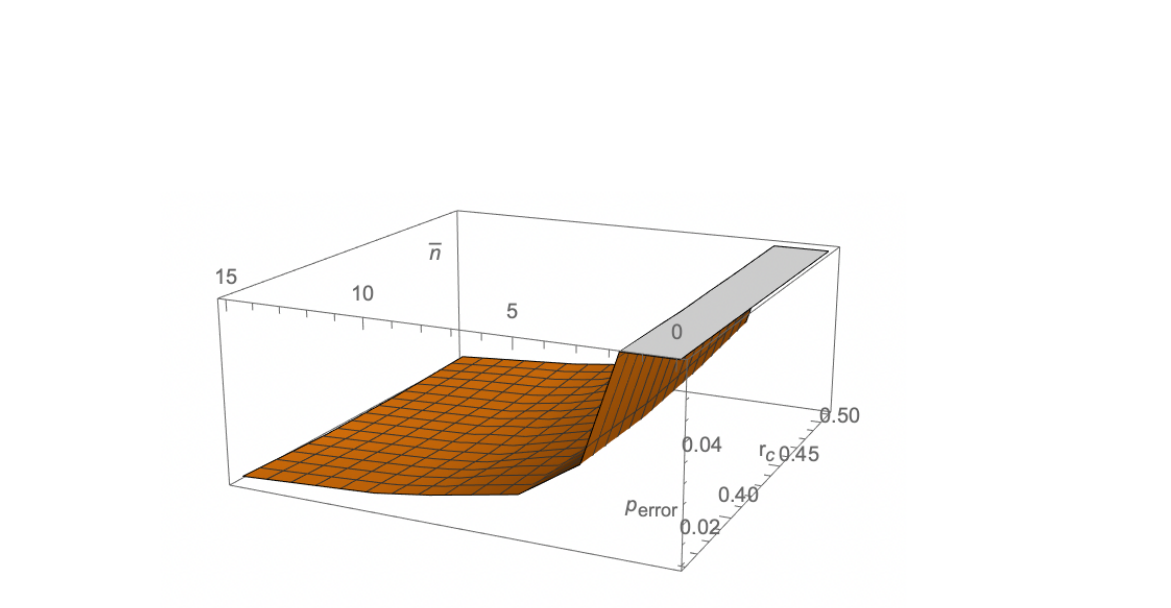}
    \includegraphics[trim = 50 270 0 200,clip, width=.8\textwidth]{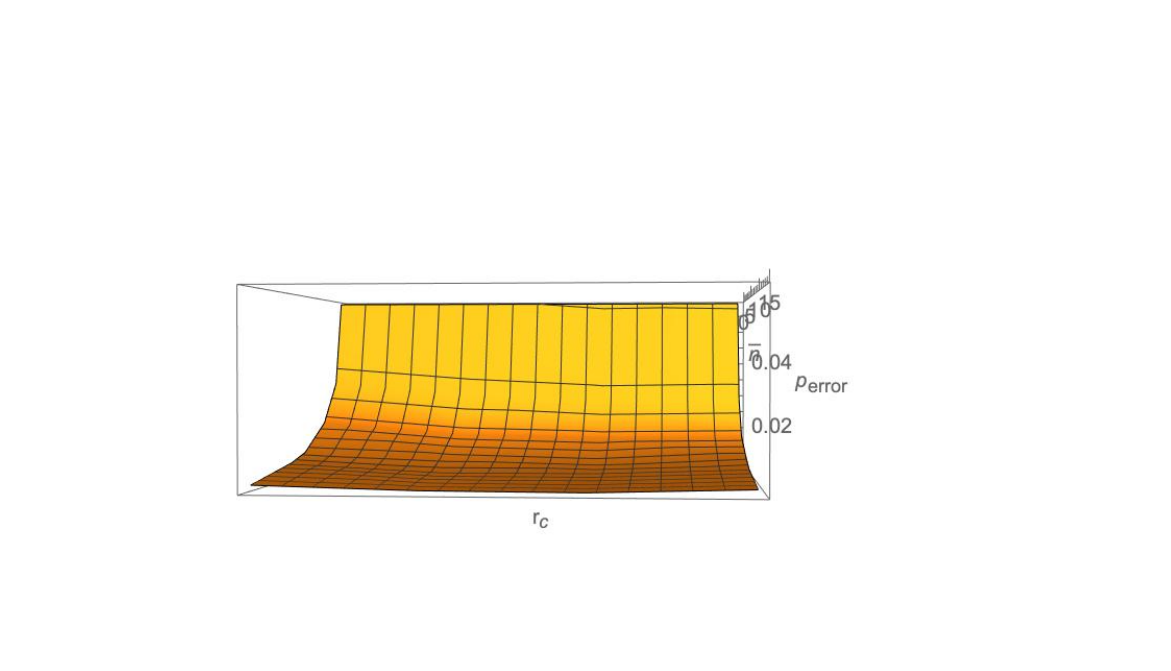}
    \caption{Two views of the plot of $p_{err}$ as a function of $r_c=- \frac{m_c}{2\beta\sqrt{\Bar{n}}}$ and $\Bar{n}$. The optimal value of $r_c$ is between .4 and .5. }
    \label{fig:perrs}
\end{figure*}

We expect that for a coherent state, the distribution of $m$ will be Gaussian \cite{OCW}. It is straightforward to compute the mean and variance of $m$ in the case where $r$ is randomly sampled from the distributions without constraint, rather than being one of two possibilities:
\begin{equation}
   \begin{aligned}
            \Bar{m}&=2\beta \Bar{r},\\
            \Bar{r}_0&=\frac{4}{\Bar{n}}\int_0^{\sqrt{\Bar{n}ln2}}r^2e^{-\frac{r^2}{\Bar{n}}}dr\approx .516\sqrt{\Bar{n}},\\
\Bar{r}_1&=-\frac{4}{\Bar{n}}\int_{-\infty}^{-\sqrt{\Bar{n}ln2}}r^2e^{-\frac{r^2}{\Bar{n}}}dr\approx -1.256\sqrt{\Bar{n}},\\
\Bar{r}_0^2&=\frac{4}{\Bar{n}}\int_0^{\sqrt{\Bar{n}ln2}}r^3e^{-\frac{r^2}{\Bar{n}}}dr\approx.307\Bar{n},\\
    \Bar{r}_1^2&=\frac{4}{\Bar{n}}\int_{-\infty}^{-\sqrt{\Bar{n}ln2}}r^3e^{-\frac{r^2}{\Bar{n}}}dr\approx 1.693\Bar{n}\\
    &\implies \Delta r_0^2\approx.041\Bar{n}\text{ ;}\Delta r_1^2\approx .131\Bar{n}.
        \end{aligned}
    \end{equation}
Likewise,
\begin{equation}
    \begin{aligned}
    \text{Var}(m)&=E(m^2)-(E(m))^2\\&=4\beta^2\Delta r^2+\beta^2+\Bar{r^2},\\
    \text{Var}(m_0)&=(.164\Bar{n}+1)\beta^2+.307\Bar{n},\\
    \text{Var}(m_1)&=(.524\Bar{n}+1)\beta^2+1.693\Bar{n}.
    \end{aligned}
\end{equation}
Sampling from a normal distribution with these parameters is simple, so we can empirically determine the optimal value of $m_c$ by simulating the transmission homodyne measurement procedure directly. The results are displayed in figure \ref{fig:perrs}.

The above refers to the case where $r$ is randomly sampled from the distributions without constraint, rather than being one of two possibilities. In latter case, where the key designates one of two specific states to be distinguished between by homodyne measurement for each transmitted bit (one state $\ket{r_0}$ from $\rho_0$ and one $\ket{r_1}$ from $\rho_1$), the optimal value calculation for $m_c\approx\beta(r_0+r_1)$ is more straightforward---it derives from optimizing the accumulated probability
\begin{equation}
    \int_{-\infty}^{m_c} \frac{e^{-\frac{(m-2\beta r_1)^2}{2(r_1^2+\beta^2)}}}{\sqrt{r_1^2+\beta^2}}-\frac{e^{-\frac{(m-2\beta r_0)^2}{2(r_0^2+\beta^2)}}}{\sqrt{r_0^2+\beta^2}}dm
\end{equation}at $m_c$---and gives the result
\begin{equation}
\begin{aligned}
    p_{\text{err}}=\frac{1}{4}\Bigg( 2+\erf\left(\frac{m_c-2\beta r_0}{\sqrt{2(\beta^2+r_0^2)}}\right)\\-\erf\left(\frac{m_c-2\beta r_1}{\sqrt{2(\beta^2+r_1^2)}}\right)\Bigg),
    \end{aligned}
    \label{homod}
\end{equation}
which does not exceed 1/2. We will compare this to the Helstrom bound once we derive a bound for Fock state communication in Section~\ref{sec:encodings}, so they can all be seen side-by-side.

\section{Particular Encoding Methods} \label{sec:encodings}

\subsection{Fock State Methods}

 Continuing from the discussion in section~\ref{sec:mappings}, if we denote any Fock states $\ket{n}$ with $n\geq 1$ as the binary 1, we must have a quantity of $n_z=\frac{N}{\Bar{n}+1}$ 0s and $N-n_z=\frac{N\Bar{n}}{\Bar{n}+1}$ 1s in the encoded string, which comes directly from the Fock state representation of $\rho_{th}$. Thus the problem is one of encoding from the set of all binary strings of length $N$ to the set of binary strings with such a ratio, of which there are $\binom{N}{n_z}$. It is straightforward to calculate the channel capacity using such an encoding, as the number of bits we can encode is simply $Nh\left(\frac{1}{\Bar{n}+1}\right)$---this provides evidence for the simplicity of the cases of extreme $\Bar{n}$ as noted in section~\ref{sec:mappings}. 

 We can approach this rate using a ``by value'' encoding. Think of an $N$-bit string as representing an integer $w$, and encode this as the $w$th smallest bit string (by value) that satisfies the above criteria for the number of 0s and 1s. We can use a theorem called the ``Christmas Stocking Theorem'' \cite{christmas} to efficiently generate the encoding, going from either the least or most significant bit. This theorem states that
\begin{equation}
    \sum_{i=0}^{k-1}\binom{n+i}{i}=\binom{k+n}{k-1}
\end{equation}
which provides a straightforward way of counting down digits. More details and examples are contained in the appendix.

\subsection{Coherent State Methods}

We want to emulate the statistics of the thermal state using coherent states sampled from our distributions, $\rho_1$ and $\rho_0$. If we have an equal number of 0s and 1s in the message, this is straightforward: we can sample from the distributions and simply transmit the result. There is a catch, however: since the coherent states are not orthogonal, there is a probability of mistaking $\rho_1$ for $\rho_0$ at the time of measurement, which produces something similar to a Pauli X error. We can protect against this by using error correction and encoded keywords; however, this requires secret key, since the encoded messages will no longer appear to be sampled from $\rho_{th}$.

For example, if we use a simple 3-bit Hamming code, there are only 2 codewords (000 and 111) and 8 possible, equally-likely bit-strings we would expect to see if sampling from $\rho_{th}$. We can make these strings appear random again by doing a bitwise XOR with a random 3-bit string, selected by generating a random number between 0 and 7. This scrambled codeword will still protect against a bit-flip error, but it requires that Bob also know the random number that was chosen. So such a scheme requires Alice and Bob to share a secret key in advance. If we wish to cut down on the amount of secret key used, we could use a shared seed for a pseudorandom number generation protocol, such as the rabbit cipher, which is thought to be cryptographically secure \cite{rabbit}. However, this would reduce the secrecy below the information-theoretic level we have been assuming up to this point.

\begin{figure*}

\includegraphics[width=.9\textwidth,clip]{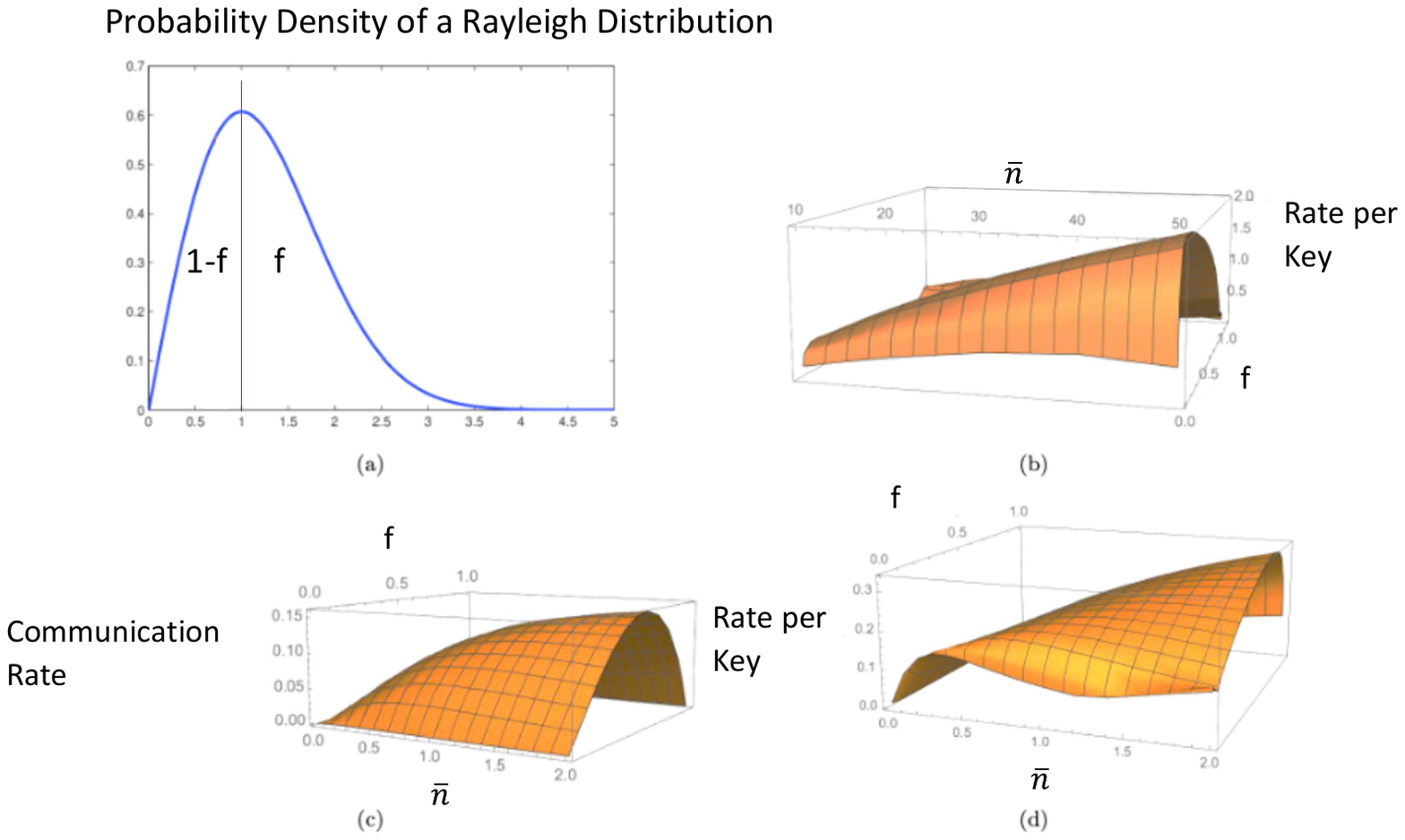}
\caption{The division of the thermal state in terms of $r$ (adapted from \cite{phdthesis}) (a) and the associated communication rates for various values of $\Bar{n}$ and $f$. In descending order, the plots display the optimized value of the communication rate per bit of secret key (b), and both the communication rate (c) and rate per bit of key (d) when the quantity to be optimized is simply the communication rate.}
\label{fig:coh}
\end{figure*}

\subsubsection{Distribution Method}

We can quantify the results of the coherent state encoding. Supposing that, instead of splitting the coherent distribution in half radially, we split it so that a fraction of the density $f$ is on the right and denotes a binary 1, and a fraction $1-f$ on the left denotes the binary 0, we can find the optimal communication rate in terms of $f$. The results in Fig. \ref{fig:coh} show that asymptotically, no secret key is needed for sufficiently large $\Bar{n}$, when we use an $f$-value closer to 0. Interestingly, there is a trade-off between the optimal communication rate in terms of minimizing $p_\text{err}$, and making the states as easy as possible to disguise. The key rate $K$ required here is given by
\begin{equation}
\begin{aligned}
    K&=h(f)-h((1-f)p(0|0)+fp(0|1))\\&+fh(p(0|1))+(1-f)h(p(0|0))\\&\geq 0
\end{aligned}
\end{equation}
where the probabilities $p(b|a)$ refer to the probability that Bob measures a state he decides is from $\rho_b$ given that Alice sent a state from $\rho_a$. We will derive expressions for these quantities below. 

\subsubsection{Pairwise Method}

We can also consider this approach from the perspective of using a finite set of states---what we called the ``pairwise'' protocol earlier. If we use a set of size $M$, with $fM$ drawn from $\rho_1$ and $(1-f)M$ drawn from $\rho_0$, we should still maintain a fidelity scaling of $F\geq 1-\frac{\Bar{n}}{2M}$, since the overall statistics are still the same as before. 

If we add factors of $1-f$ in front of $p(1|0)$ and $f$ in front of $p(0|0)$ in the expression we previously derived in \ref{homod} for $p_{\text{err}}$ under homodyne measurement, with these weightings we now have
\begin{equation}
    m_c\approx \beta(r_0+r_1)+\frac{\beta\log(\frac{1}{f}-1)}{2(r_0-r_1)}
\end{equation}
and
\begin{equation}
\begin{aligned}
    p_{\text{err}}=\frac{1}{4} \Bigg(f \left(\text{erf}\left(\frac{m_c-2\beta r_0}{\sqrt{2\beta ^2+r_0^2}}\right)+1\right)\\-(1-f)
   \left(\text{erf}\left(\frac{m_c-2\beta r_1}{\sqrt{2\beta ^2+r_1^2}}\right)+1\right)\Bigg).
   \end{aligned}
\end{equation}
We see that for $f=1/2$ the value of $m_c$ reverts to $\beta(r_0+r_1)$.

For optimal generalized (Helstrom) measurements, we consider the quantity
\begin{equation}
    M=(1-f)\ket{\alpha_0}\bra{\alpha_0}-f\ket{\alpha_1}\bra{\alpha_1}
\end{equation}
If we orthogonalize the basis using $\eta=\bra{\alpha_0}\ket{\alpha_1}\in\mathbb{R}$, we can express this as
\begin{equation}
\begin{aligned}
M=(1-f-f|\eta^2|)\ket{\alpha_0}\bra{\alpha_0}-f(1-|\eta|^2)\ket{\alpha_0^\perp}\bra{\alpha_0^\perp}\\-f\eta\sqrt{1-|\eta|^2}\ket{\alpha_0}\bra{\alpha_0^\perp}-f\eta\sqrt{1-|\eta|^2}\ket{\alpha_0^\perp}\bra{\alpha_0}
\end{aligned}
\end{equation}
with
\begin{equation}
    \ket{\alpha_1}=\eta\ket{\alpha_0}+\sqrt{1-\eta^2}\ket{\alpha_0^\perp}.
\end{equation}
Then we have that in this basis
\begin{equation}
    \begin{aligned}
    p(0|0)&=\text{Tr}\Big(\ket{\alpha_0}\braket{\alpha_0|v_0}\bra{v_0}\Big)=|\braket{\alpha_0|v_0}|^2
    \\
    p(0|1)&=\text{Tr}\Big(
    % \begin{pmatrix}\beta \\\sqrt{1-\beta^2}
    % \end{pmatrix}\begin{pmatrix}\beta &
    % \sqrt{1-\beta^2}
    % \end{pmatrix}
    \ket{\alpha_1}\braket{\alpha_1|v_0}\bra{v_0}\Big)=|\braket{\alpha_1|v_0}|^2
    \end{aligned}
\end{equation}
where $v_i$ is the eigenvector of $M$ corresponding to the $\alpha_i$ eigenspace.

\begin{figure*}
    \centering
    \includegraphics[trim = 50 300 0 200,clip,width=.8\textwidth]{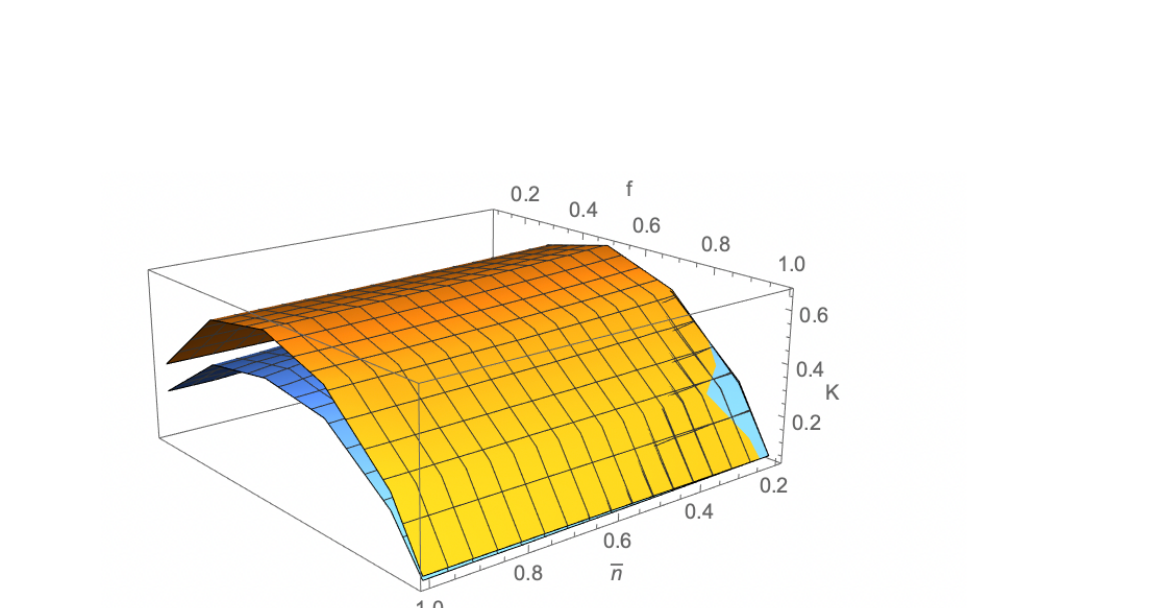}
    \includegraphics[trim = 50 300 0 200,clip,width=.8\textwidth]{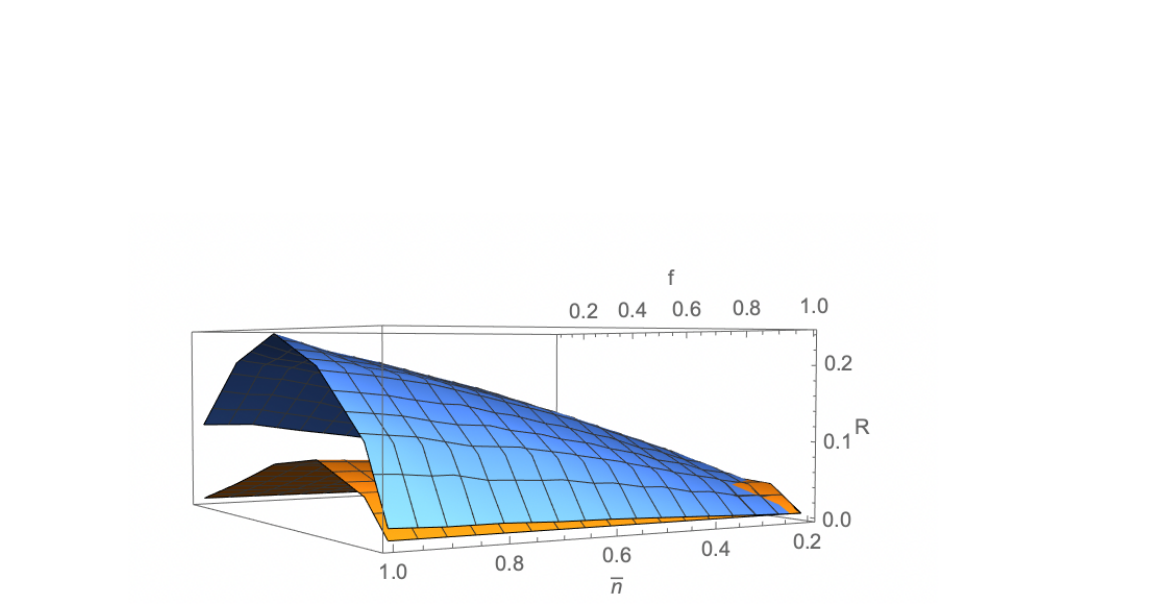}
    \includegraphics[trim = 50 270 0 200,clip,width=.8\textwidth]{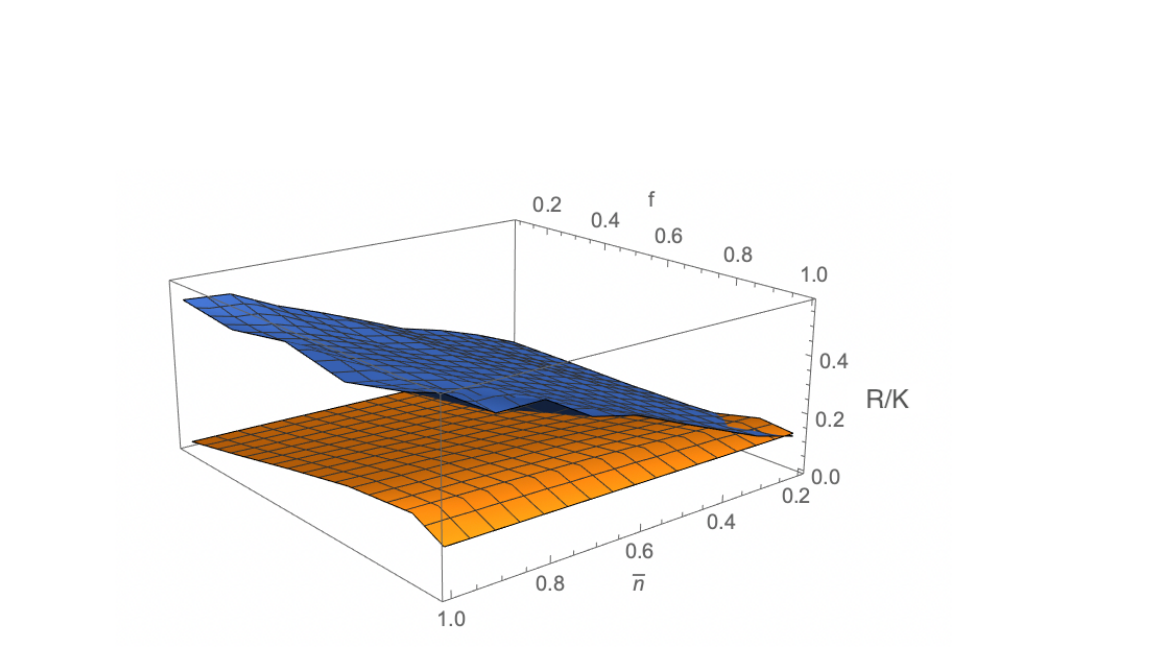}
    \caption{The communication rate $R$, key rate $K$, and quotient $\frac{R}{K}$ for encoded transmission using homodyne-type (orange) and Helstrom-type (blue) measurements at a variety of $f$ and $\Bar{n}$ values.}
    \label{fig:finalcomp}
\end{figure*}

From these quantities we can roughly determine the rate of this communication method, which is limited by the entropy difference between the encoded and decoded information:
\begin{equation}
    R\approx h(q)-h(q|x_i)
\end{equation}
and the key rate is set by the entropy difference between a simulated binary symmetric channel and the mutual information of the actual channel
\begin{equation}
    K=h(f)-R
\end{equation}
where $q=(1-f)p(0|0)+fp(0|1)=\sum_i p_i p(0|x_i)$. Then, we can plot the associated quantities, such as $K$, $R$, and $\frac{R}{K}$, and we see in Fig.~\ref{fig:finalcomp} that the ratio is fairly small for the homodyne measurement case but not for the Helstrom case.

We will also need additional shared secret key to specify which pair of states is being used for each transmitted bit. To specify a pair of states, one for each bit value, requires an additional $\log((1-f)(f)M^2)$ bits of key. The communication rate, however, remains the same as described above. Note that the asymptotic communication rate per bit of key can be demonstrated analytically to exceed 1 by examining the limiting behavior of the expression for $\frac{R}{K}$. 

Note that the value of $\Bar{n}$ manifests itself in the specific coherent states that will be drawn from the distributions, rather than being directly visible in the binary transmitted bits as in the Fock case. If the message does not contain a roughly equal number of 1s and 0s, we can use an encoding, such as the one described in the appendix, to compensate for that. However, in that case the message should first be compressed, which increases the entropy of the transmitted string; and if need be, a sublinear amount of secret key can be used to make the message string indistinguishable from a purely random bit string \cite{noichris,noilchris}.

\begin{figure*}
    \centering
    \includegraphics[trim = 50 270 0 200,clip, width=1.1\textwidth]{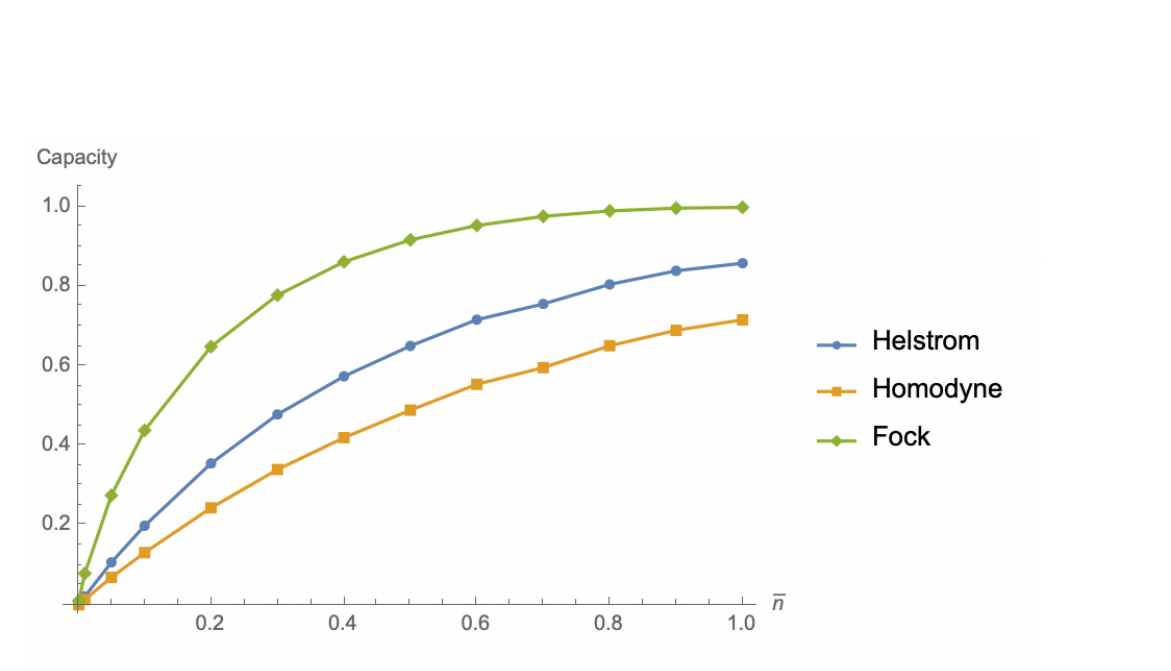}
    \caption{A comparison of the communication rates using the Fock and coherent state encodings. In the Fock case this results from the theoretical optimum measurement based on the binary symmetric channel capacity and in the latter from optimal homodyne measurements and generalized measurements that approach the Helstrom bound. In the coherent state cases, the results are derived from sampling from the constituent distributions, since an analytical result is not as straightforward to obtain.}
    \label{fig:cohfockcomp}
\end{figure*}

\section{Discussion and Future Work} \label{sec:discussion}

As we can see, the Fock state encoding is superior in the noiseless case and also doesn't require key. Moreover, it is straightforward and the encoding has a clean visual representation based on Pascal's triangle (as shown in the Appendix). However, coherent states are more resistant to noise and are easier to generate in the lab, although they perform noticeably worse under either ideal (Helstrom limit) or homodyne measurements. In both cases, however, it is viable to communicate information covertly using the above-described methods for any channel parameter $\Bar{n}$, and even in the coherent case the amount of secret key required is low compared to, for example, a one-time pad.

There are many possible future directions to explore to build on this work. An interesting problem is that of modeling noisy channels---for example, one with an existing thermal noise background---and the transmission of coherent states through such a channel (which is a well-understood problem) \cite{Vourdas}. Such a study would provide a more thorough grounding for this work's study of coherent state methods, which are suboptimal in the noiseless case we have examined.
    
Another important problem is that of transmitting {\it quantum} information, including entanglement. In this work we have only considered classical information transmission. It is unclear what kinds of encodings can be used for quantum information and how well they preserve the entanglement of the system, both in the noisy and the noiseless case. Such a work might also delve into the potential applications of this communication method to teleportation and superdense coding protocols, for example, and methods of making those more secure under the type of schemes covered in this paper.
    
A final promising area of study could be different key utilization protocols, with the goal of utilizing the inherent noise protection assumptions of steganography to efficiently scale the encryption process, and using other quantum communications as a vehicle for encoding steganographic information. This aims to get around this work's requirement that secret key is needed for non-orthogonal state discrimination by exploiting the information difference between Alice, Bob, and Eve to communicate a secret key seed, without compromising the communication rate derived above.

\section*{Acknowledgements} \label{sec:acknowledgements}
The authors would like to thank Jonathan Habif and Haley Weinstein for helpful discussions and insights. This research was supported in part by NSF Grants 1719778 and 1911089.

\appendix*
\section{} \label{sec:appendix}
    \subsection{Proof of Eq.~(\ref{equation:fidelity})}
    We start from the fidelity between 
    \begin{equation}
        \rho_c=\rho=\frac{1}{M}\sum_{j=1}^Me^{-r_j^2}\sum_n\frac{r_j^{2n}}{n!}\ket{n}\bra{n}
    \end{equation}
    and the thermal state $\rho_{th}$:
    \begin{equation}
        \sqrt{F(\rho, \rho_th)}=\text{Tr}\sqrt{\rho_{th}^{1/2}\rho\rho_{th}^{1/2}}.
    \end{equation}
    If we define the quantity $\Delta \rho=\rho-\rho_{th}$, then we know immediately that $E(\Delta\rho)=0$ since $E(\rho)=\rho_{th}$. Then since $[\rho_{th}, \rho]=0$, we can write
    \begin{equation}
        \sqrt{F}=\text{Tr}\left(\rho_{th}\sqrt{I+\rho_{th}^{-1}\Delta\rho}\right)\
    \end{equation}
    which can be lower-bounded by taking the binomial expansion of the square root. We can then take the ensemble average:
    \begin{equation}
        \begin{aligned}
        \label{eqn:exp}
            E \Big( &\text{Tr}\Big(\rho_{th}\Big(I+\frac{1}{2}\rho_{th}^{-1}\Delta\rho-\frac{1}{2}(\rho_{th}^{-1}\Delta \rho)^2\Big)\Big)\Big)\\
            &=1-\frac{1}{2}\text{Tr}(\rho_{th}^{-1}E(\Delta\rho^2))\\
            &=\frac{3}{2}-\frac{1}{2}\text{Tr}(\rho_{th}^{-1}E(\rho^2))
        \end{aligned}
    \end{equation}
    where here \begin{equation}
        \rho^2=\left(\frac{1}{M}\sum_{j=1}^M e^{-r_j^2}\sum_n\frac{r_j^{2n}}{n!}\ket{n}\bra{n}\right)^2=\frac{1}{M^2}\sum_{jk}\rho_j\rho_k.
    \end{equation}
    We have that the $\rho_j$ are independent and that for each one $E(\rho_j)=\rho_{th}$, so if we consider the cases only where all $r_j$ are equal, which we'll call $\rho_1$, we have
    \begin{equation}
        E(\rho^2)=\frac{M^2-M}{M}\rho_{th}^2+\frac{1}{M}E(\rho_1^2)
    \end{equation}
    which makes the result of Eq. ~(\ref{eqn:exp})
    \begin{equation}
        \sqrt{F}\geq 1+\frac{1}{2M}-\frac{1}{2M}\text{Tr}(\rho_{th}^{-1}E(\rho_1^2)).
    \end{equation}
    We can now finally evaluate this trace term:
    \begin{equation}
        \begin{aligned}
        &\text{Tr}(\rho_{th}^{-1}E(\rho_1^2))\\
            &=(\Bar{n}+1)E\left(e^{-2r^2}\sum_{n=0}^\infty \left(\frac{\Bar{n}+1}{\Bar{n}}\right)^n\frac{r^{4n}}{(2n)!^2}\right)\\&=\sum_{n=0}^\infty\frac{2}{(n!)^2}\left(\frac{\Bar{n}+1}{\Bar{n}}\right)^{n+1}\int_0^\infty e^{-(2+1/\Bar{n})r^2}r^{4n}dr\\
            &=\sum_{n=0}^\infty\frac{(2n)!}{(n!)^2}\frac{\Bar{n}+1}{2\Bar{n}+1}\left(\frac{\Bar{n}+1}{\Bar{n}(2+1/\Bar{n})^2}\right)^n\\
            &=\frac{1}{2\pi}\frac{\Bar{n}+1}{2\Bar{n}+1}\sum_{n=0}^\infty \left(\frac{\Bar{n}+1}{\Bar{n}(2+1/\Bar{n})^2}\right)^n\int_0^{2\pi} (4\text{cos}^2\phi)^nd\phi\\
            &=\frac{1}{2\pi}\frac{\Bar{n}+1}{2\Bar{n}+1}\int_0^{2\pi}\sum_{n=0}^\infty \left(\frac{(\Bar{n}+1)4\text{cos}^2\phi}{\Bar{n}(2+1/\Bar{n})^2}\right)^nd\phi\\
            &=\Bar{n}+1
        \end{aligned}
    \end{equation}
    where we have used that
    \begin{equation}
        \frac{(2n)!}{(n!)^2}=\frac{1}{2\pi}\int_0^{2\pi} (4\text{cos}^2\phi)^nd\phi.
    \end{equation}
    This makes the final result of Eq. ~(\ref{eqn:exp})
    \begin{equation}
        \sqrt{F}\geq 1-\frac{\Bar{n}}{2M}
    \end{equation}
    in the average case.

\subsection{Discretizing the Circle}

The second bound is when the circle is discretized over $\theta$. This encoding gives us that for $j=1...M$ and $k=0...L-1$ and $N=ML$
\begin{equation}
    \ket{\alpha_{jk}}=\ket{r_je^{\frac{2\pi ik}{L}}}=\sum_{jk}\sum_{n=0}^\infty e^{-r_j^2/2}\frac{(r_je^{2\pi ik/L})^n}{\sqrt{n!}}\ket{n} .
\end{equation}
Then we have
\begin{equation}
    \begin{aligned}
        \rho_c&=\frac{1}{N}\sum_{n, n'} \sum_{jk} e^{-r_j^2}\frac{r_j^{n+n'}}{\sqrt{n!n'!}}e^{2\pi ik(n-n')/L}\ket{n}\bra{n'}\\
        &=\frac{1}{N}\sum_{n, n'} \sum_{j} e^{-r_j^2}\frac{r_j^{n+n'}}{\sqrt{n!n'!}}\frac{1-e^{2\pi i(n-n')}}{1-e^{2\pi i(n-n')/L}}\ket{n}\bra{n'}\\
        &=\frac{1}{N}\sum_{n=0}^\infty \sum_j e^{-r_j^2}\frac{r_j^{2n}}{n!}\ket{n}\bra{n}
    \end{aligned}
\end{equation}
since the geometric series sums to $\delta_{nn'}$. We can now evaluate the fidelity:
\begin{equation}
    \begin{aligned}
        &\sqrt{F}
        \\&=\text{Tr}\left(\sqrt{\sum_{n=0}^\infty\frac{1}{N(\Bar{n}+1)}\left(\frac{\Bar{n}}{\Bar{n}+1}\right)^n\frac{1}{n!}\sum_je^{-r_j^2}r_j^{2n}\ket{n}\bra{n}}\right).
    \end{aligned}
\end{equation}
This gives the same bound as before:
\begin{equation}
    \sqrt{F}\geq 1-\frac{\Bar{n}}{2N}
\end{equation}
If we instead consider the distributions $\rho_1$ and $\rho_0$, we can easily show, at least, that $\sqrt{F}\to 1$ as $N\to\infty$ (since the integrals over the distributions are hard to compute) starting from Eq. ~(\ref{eqn:exp}):
\begin{equation}
\begin{aligned}
\sqrt{F}&=1+\frac{1}{2N}-\frac{1}{2N}\text{Tr}\left(\rho_{th}^{-1}E[\rho_i^2]\right)\\
E[\rho_i]&=\frac{2}{\Bar{n}}\Big(\int_0^{r_{1/2}} r^{4n+1}e^{-(2+1/\Bar{n})r^2}dr\\&+\int_{-r_{1/2}}^{-\infty} r^{4n+1}e^{-(2+1/\Bar{n})r^2}dr\Big)\\&\leq \frac{2}{\Bar{n}}\int_0^\infty r^{4n+1}e^{-(2+1/\Bar{n})r^2}dr\\
&\implies -\Bar{n}-1\leq \text{Tr}(\rho_{th}^{-1}E[\rho_i])\leq \Bar{n}+1\\
&\implies \sqrt{F}\geq 1-\frac{\Bar{n}}{2N}.
\end{aligned}
\end{equation}
    
\subsection{Derivation of Vertical Angle Bound with no Key}

We start with the two states, given $c=1+1/\Bar{n}$:
\begin{equation}
\begin{aligned}
\rho_0&=\frac{4}{\Bar{n}}\int_0^{r_{1/2}}\sum_{n=0}^\infty e^{-(1+1/\Bar{n})r^2}\frac{r^{2n+1}}{n!}\ket{n}\bra{n}\\
&=\frac{2^{-\Bar{n}}}{\Bar{n}+1}\sum_{n=0}^\infty \frac{1}{c^n}\left(\sum_{k=0}^n\frac{(cr_{1/2}^2)^k}{k!}\right)\ket{n}\bra{n}
\end{aligned}
\end{equation}
and
\begin{equation}
\begin{aligned}
\rho_1&=\frac{4}{\Bar{n}}\int_{r_{1/2}}^\infty\sum_{n=0}^\infty e^{-(1+1/\Bar{n})r^2}\frac{r^{2n+1}}{n!}\ket{n}\bra{n}\\
&=\frac{2^{-\Bar{n}}}{\Bar{n}+1}\sum_{n=0}^\infty \frac{1}{c^n}\left(\sum_{k=n+1}^\infty\frac{(cr_{1/2}^2)^k}{k!}\right)\ket{n}\bra{n}.
\end{aligned}
\end{equation}

The fidelity between these two states is given by
\begin{equation}
\sqrt{F}=\text{Tr}\sqrt{\rho_0\rho_1}=\frac{2}{\Bar{n}+1}\sum_{n=0}^\infty\left(\frac{\Bar{n}}{\Bar{n}+1}\right)^n\sqrt{Q_n(1-Q_n)}
\end{equation}
where $Q_n=2^{-(\Bar{n}+1)}\sum_{k=0}^n\frac{(cr_{1/2}^2)^k}{k!}$ is the $n$th cumulant of the Poisson process with parameter $\lambda=cr_{1/2}^2$.

We can rewrite the trace distance as 
\begin{equation}
\begin{aligned}
\frac{1}{2}||\rho_0-\rho_1||=\frac{1}{\Bar{n}+1}\Big(\sum_{n=0}^{N_{1/2}}\left(\frac{\Bar{n}}{\Bar{n}+1}\right)^n(1-2Q_n)\\+\sum_{n=N_{1/2}}^\infty\left(\frac{\Bar{n}}{\Bar{n}+1}\right)^n(2Q_n-1)\Big)
\end{aligned}
\end{equation}
where $N_{1/2}$ is such that $Q_n<1/2$ iff $n<N_{1/2}$.

We take each term in the sum in turn:
\begin{equation}
\begin{aligned}
&\frac{1}{\Bar{n}+1}\left(\sum_{n=0}^{N_{1/2}}\left(\frac{\Bar{n}}{\Bar{n}+1}\right)^n(1-2Q_n)\right)\\&=\left(1-\left(\frac{\Bar{n}}{\Bar{n}+1}\right)^{N_{1/2}+1}\right)\\&-\frac{2^{-\Bar{n}}}{\Bar{n}+1}\sum_{k=0}^{N_{1/2}}\frac{((\Bar{n}+1)\ln 2))^k}{k!}\left(\frac{(\frac{\Bar{n}}{\Bar{n}+1})^k-\frac{\Bar{n}}{\Bar{n}+1})^{N_{1/2}+1}}{1-\frac{\Bar{n}}{\Bar{n}+1}}\right)\\
&=1+(2Q_{N_{1/2}}-1)\left(\frac{\Bar{n}}{\Bar{n}+1}\right)^{N_{1/2}+1}-\tilde{Q}_{N_{1/2}},
\end{aligned}
\end{equation}
where $\Tilde{Q}_{n}$ is the CDF for the Poisson process with $\lambda=\Bar{n}\ln 2$.

The second sum is
\begin{equation}
\begin{aligned}
&\frac{1}{\Bar{n}+1}\sum_{k=N_{1/2}+1}^\infty\left(\frac{\Bar{n}}{\Bar{n}+1}\right)^n(2Q_n-1)
\\
&=\sum_{n=N_{1/2}+1}^{\infty}\left(\frac{\Bar{n}}{\Bar{n}+1}\right)^n\sum_{k=0}^n\frac{((\Bar{n}+1)\ln2)^k}{k!} \\&-\left(\frac{\Bar{n}}{\Bar{n}+1}\right)^{N_{1/2}+1}
\\
&=-\left(\frac{\Bar{n}}{\Bar{n}+1}\right)^{N_{1/2}+1}\\&+\frac{2^{-\Bar{n}}}{\Bar{n}+1}\Big(\sum_{k=0}^{N_{1/2}}\frac{((\Bar{n}+1)\ln2)^k}{k!}\sum_{n=N_{1/2}+1}^\infty \left(\frac{\Bar{n}}{\Bar{n}+1}\right)^n\\&+\sum_{k=N_{1/2}+1}^{\infty}\frac{((\Bar{n}+1)\ln2)^k}{k!}\sum_{n=k}^\infty \left(\frac{\Bar{n}}{\Bar{n}+1}\right)^n\Big)\\&=1+(2Q_{N_{1/2}}-1)\left(\frac{\Bar{n}}{\Bar{n}+1}\right)^{N_{1/2}+1}-\tilde{Q}_{N_{1/2}}.
\end{aligned}
\end{equation}

Combining these sums gives us the final result:
\begin{equation}
\begin{aligned}
  \frac{1}{2}\left|\left|\rho_0-\rho_1\right|\right|&=2\left(1+(2Q_{N_{\frac{1}{2}}}-1)\left(\frac{\Bar{n}}{\Bar{n}+1}\right)^{N_\frac{1}{2}+1}-\Tilde{Q}_{N_\frac{1}{2}}\right)
\end{aligned}
\end{equation}

\begin{figure}
    \centering
    \includegraphics[scale=.3]{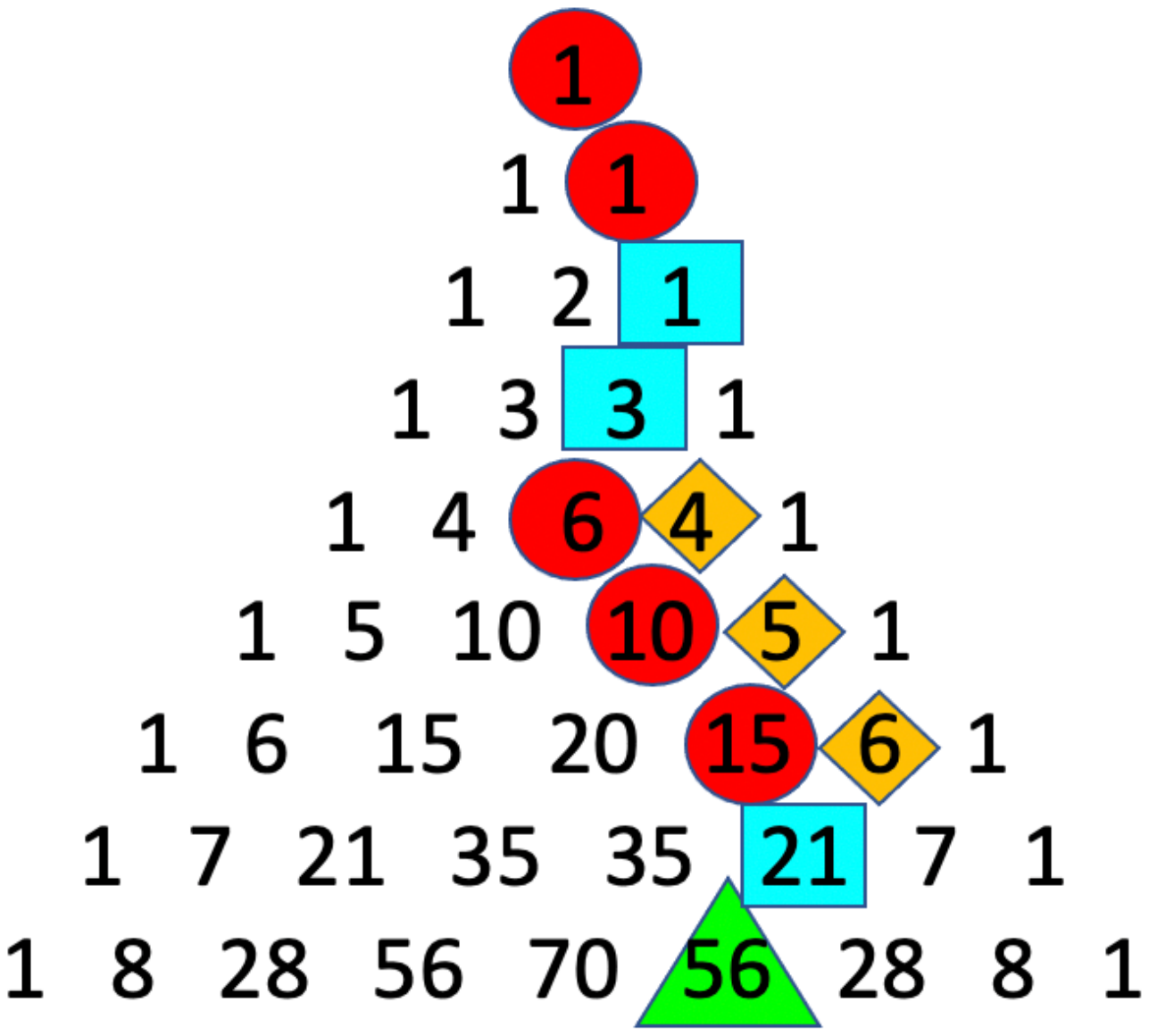}
    \caption{This figure illustrates the process Alice uses to construct the encoded message as described above on Pascal's Triangle. First, she calculates that the message is transmissible using 8 bits, since 41<56 (green), as above. Then, she determines the message (41) is in the largest 21 numbers, and so she boxes 21 and the first digit is 1 so she moves right. Then she determines 41 is neither in the largest 6, nor 6+5, nor 6+5+4 messages, and so circles the numbers 6, 10, and 15, and moves left each time, with each circle signifying a zero (the intermediate numbers 4, 5, and 6 are also highlighted since it is useful to keep track of their values). These steps are repeated until Alice reaches the right edge of the triangle, at which point she knows all remaining digits are 0. As this figure shows, this is an efficient encoding, as only a linear amount of combinations in the starting position need to be computed.}
    \label{fig:christmas}
\end{figure}

\subsection{Encoding Method for Fock States}
Elaborating on the results of Section V, we can imagine a particular encoding method for Fock state communication. Let's consider a particular example. Suppose $\Bar{n}=.56$ and Alice wants to communicate the six-bit string 101001, which we can think of as the binary expression for the number 41. (Note that this protocol requires Bob to know the size of the message being communicated). If Alice wants to know how many bits she needs to encode an $x$-bit number with $\frac{M}{\Bar{n}+1}$ zeroes, she solves the equation \begin{equation}
    \binom{M}{\frac{M}{\Bar{n}+1}}=2^{x+1}-1.
\end{equation}
Alice can transmit her string with $M=8$ bits: the condition is that \begin{equation}
    \binom{M}{\frac{M}{\Bar{n}+1}}=2^{Mh(\frac{1}{\Bar{n}+1})}\geq |m| ,
\end{equation}
where we use $|m|$ to denote the value of the message, which is 41. Therefore, Alice's message should consist of 5 bits of 0, 3 bits of 1, and we can verify that $\binom{8}{3}=56>41$. 

As we will show, 41 is encoded by the binary string 10001100, which is the binary expression for 140, the $41st$-smallest number with an 8 bit string representation (possibly with leading zeroes) with the appropriate ratio of 1s to 0s. To find this string,  we follow a process derived from the ``Christmas Stocking Theorem'' mentioned above, which we detail in figure \ref{fig:christmas}. First, consider that if we are using $N$ bits, $n_z$ of which are 0, there are $\binom{N-1}{n_z}$ strings with 1 in the first spot. There are $\binom{N-2}{n_z-1}$ strings with 0 in the first spot and 1 in the second spot. And so on. Alice can subtract these from $56$ total possible strings to find the string in question. For example, the first $\binom{7}{5}=21$ bits have a 1 at the start, and $41>56-21$ so Alice's message must start with 1, denoted by the blue square over the 21 in the figure. Then there are $\binom{6}{5}=6$ strings with 11, so the second bit she sends must be 0, since $41\leq56-6$ (denoted by the red square over the 6 in the figure). $56-6-5-4=41$ so the string so far is 10001. Actually, at this point we have reached the $41st$ largest string so Alice's message is the largest such string with that prefix, 10001100=140, and the remaining digits are in red in the figure since they correspond to the trailing 0s now that we have allocated all the 1s.

Once Alice knows that this is the encoded string she wants to send, she generates 8 Fock states in a ratio of 5:3 zeroes to ones (where ``one'' here refers to a mode greater than 0, with the appropriate statistics to emulate the thermal state). Then she sends one of the $\ket{1+}$ states for every position corresponding to 1 in the string 10001100 and a $\ket{0}$ state for each 0. Bob measures the Fock states and should receive 1001100 as the most likely string, after inverting the above algorithm, which is straightforward. Eve should see something that looks like a thermal state: after all, it is bitwise random and has the required overall statistics.

\subsection{Practical Fidelity Bounds for the Fock Encoding}

In practice, the Fock state encoding is not exactly equivalent to a thermal state. While we would like to send the state
\begin{equation}
    \rho = \frac{1}{\Bar{n}+1}\ket{\Bar{0}}\bra{\Bar{0}}+\frac{\Bar{n}}{\Bar{n}+1}\ket{\Bar{1}}\bra{\Bar{1}} ,
\end{equation}
in practice we have a finite number of bits, so we are sending either the state
\begin{equation}
    \rho'=\frac{1}{N}\Big(\Big\lfloor\frac{N}{\Bar{n}+1}\Big\rfloor\ket{\Bar{0}}\bra{\Bar{0}}+\Big\lceil\frac{N\Bar{n}}{\Bar{n}+1}\Big\rceil\ket{\Bar{1}}\Bra{\Bar{1}}\Big) ,
\end{equation}
or the state
\begin{equation}
    \rho''=\frac{1}{N}\Big(\Big\lceil\frac{N}{\Bar{n}+1}\Big\rceil\ket{\Bar{0}}\bra{\Bar{0}}+\Big\lfloor\frac{N\Bar{n}}{\Bar{n}+1}\Big\rfloor\ket{\Bar{1}}\bra{\Bar{1}}\Big) ,
\end{equation}
(whichever maximizes the fidelity). Another way of writing this is, e.g.,
\begin{equation}
\begin{aligned}
     \rho'=&\frac{\left\lfloor\frac{N}{\Bar{n}+1}\right\rfloor}{N}\ket{0}\bra{0}\\&+\frac{1}{N(\Bar{n}+1)}\sum_{n=1}^\infty \left\lceil\frac{N\Bar{n}}{\Bar{n}+1}\right\rceil \left(\frac{\Bar{n}}{\Bar{n}+1}\right)^{n-1}\ket{n}\bra{n} .
\end{aligned}
\end{equation}

We would like to compute, as a function of $\Bar{n}$,
\begin{equation}
\begin{aligned}
    &\text{max}_{\sigma\in\{\rho', \rho''\}} F(\rho, \sigma)\\
    &=\text{max}(\text{Tr}(\sqrt{\rho\sigma}))\\
    &=\text{max}\left(\text{Tr}\sqrt{\sum_n\frac{1}{\Bar{n}+1}\left(\frac{\Bar{n}}{\Bar{n}+1}\right)^nc_n\ket{n}\bra{n}}\right)\\
    &=\text{e.g. }\sqrt{\frac{\lfloor\frac{N}{\Bar{n}+1}\rfloor} {N(\Bar{n}+1)}}+\sum_{n=1}^\infty\frac{1}{\Bar{n}+1} \sqrt{\frac{\lceil\frac{N\Bar{n}}{\Bar{n}+1}\rceil\Bar{n}} {N(\Bar{n}+1)}}\left(\frac{\Bar{n}}{\Bar{n}+1}\right)^n\\
&\geq 1-\frac{1}{8}(\frac{(1-\frac{\lfloor\frac{N}{\Bar{n}+1}\rfloor(\Bar{n}+1)}{N})^2}{\Bar{n}+1}+\frac{\Bar{n}(1-\frac{\lceil\frac{N\Bar{n}}{\Bar{n}+1}\rceil(\frac{\Bar{n}+1}{\Bar{n}})}{N})^2}{\Bar{n}+1})-\frac{\left(\frac{\Bar{n}+1}{\Bar{n}}\right)^2}{16N^3}\\&=1-\frac{1}{8}(\frac{\left\lceil\frac{N\Bar{n}}{\Bar{n}+1}\right\rceil^2\frac{\Bar{n}+1}{\Bar{n}}+\left\lfloor\frac{N}{\Bar{n}+1}\right\rfloor^2(\Bar{n}+1)}{N^2}-1)-\frac{\left(\frac{\Bar{n}+1}{\Bar{n}}\right)^2}{16N^3}\\
&\geq 1-\frac{1}{8}\left(\frac{(\frac{N\Bar{n}}{\Bar{n}+1}+1)^2(\frac{\Bar{n}+1}{\Bar{n}})+\left(\frac{N}{\Bar{n}+1}\right)^2(\Bar{n}+1)}{N^2}-1\right)-\frac{\left(\frac{\Bar{n}+1}{\Bar{n}}\right)^2}{16N^3}\\&=1-\frac{1}{8}\left(\frac{2N+1}{N^2}+\frac{1}{N^2\Bar{n}}\right)-\frac{1}{16}\frac{\left(\frac{\Bar{n}+1}{\Bar{n}}\right)^2}{N^3}, \Bar{n}\geq1,\\
&\geq 1-\frac{1}{8}\left(\frac{2N+\Bar{n}+1}{N^2}\right)-\frac{1}{16}\frac{(\Bar{n}+1)^2}{N^3} , \Bar{n}<1 ,  \end{aligned}
\end{equation}
%If we wish to be more precise, we can repeat this procedure while enforcing the ceiling and floor function for all terms, but this complicates the problem significantly without better addressing the issue at hand, and is likely not an improvement on this bound. The expression would then look something like
%\begin{equation}
%    \rho' = \frac{1}{N}\left(\left\lfloor \frac{N}{\Bar{n}+1}\right\rfloor\ket{0}\bra{0}+\frac{1}{\Bar{n}+1}\sum_{n=1}^\infty \left\lceil N\left(\frac{\Bar{n}}{\Bar{n}+1}\right)^n\right\rceil\ket{n}\bra{n}\right)
%\end{equation}
%except each term can be either a floor or a ceiling.
where the third order correction is due to Taylor's theorem. The above chain of inequalities shows that when $N$ is fairly large, the fidelity between the encoded state and the thermal state is very close to 1. At the cost of sharing some additional secret key, Alice and Bob can actually make the fidelity (almost) perfect. The gap in fidelity between the encoded state and the thermal state results from using integer approximations to $N/(\bar{n}+1)$ and $N\bar{n}/(\bar{n}+1)$. But Alice and Bob can, in principle, randomly choose the numbers of 0s and 1s to be used in the encoding, from the binomial distribution with probabilities corresponding to the thermal state: $p(0)=1/(\bar{n}+1)$ and $p(1)= \bar{n}/(\bar{n}+1)$. By averaging over this choice, one can match the thermal state perfectly.

\end{document}